\documentclass[twoside]{ilcws08}
\usepackage[latin1]{inputenc}
\usepackage[dvips]{graphicx,epsfig,color}
\usepackage{wrapfig,rotating}
\usepackage{amssymb,amsmath,array}

\begin{document}
\title{
Beam Size Measurement with Pair Monitor and BeamCal } 
\author{Kazutoshi Ito$^1$, Akiya Miyamoto$^2$, Tadashi Nagamine$^1$,
Yutaro Sato$^1$,
Yosuke Takubo$^1$,\\
 Toshiaki Tauchi$^2$ and Hitoshi Yamamoto$^1$
\vspace{.3cm}\\
1- Department of Physics, Tohoku University \\
Research Center for Neutrino Science, Tohoku University Sendai , Japan
\vspace{.1cm}\\
2- High Energy Accelerator Research Organization (KEK), Tsukuba, Japan
}

\maketitle

\begin{abstract}
 At the International Linear Collider (ILC),
 measurement of the beam profile at the interaction point (IP)
 is a key issue to achieve high luminosity.
 We studied the beam size measurement by combination of
 Pair Monitor and BeamCal.
 We obtained measurement accuracies of 2.7\%, and 6.6\%
for the horizontal ($\sigma_x$), and vertical ($\sigma_y$) beam sizes,
 respectively, for 50 bunch crossings.
\end{abstract}

\section{Introduction}
\label{introduction}

In the ILC, measurement of the beam size at IP is important
to keep high luminosity because the luminosity critically
depends on beam size.
Since the vertical beam size is very small (5.7 nm),
it must be measured with in 1 nm accuracy. 
We have studied measurement of the beam profile with Pair Monitor
\cite{url, kazuto} and
BeamCal.
Since the beam energy and the particle density is high at ILC, 
a large number of $e^+e^-$ pairs
are created during the bunch crossing by three
incoherent processes; Breit-Wheeler process, Bethe-Heitler process
and Landau-Lifshitz process. The generated $e^\pm$ pairs are usually 
referred to as the pair background.
The particles with the same charge as the oncoming beam
are scattered with large angles and carry information on the beam
profile at IP\cite{TauchiAndYokoya}.

Pair Monitor and BeamCal offer the independent information, and
combining them
results in better measurement accuracies.
In this paper, we report a reconstruction of the beam sizes
($\sigma_x, \sigma_y$) using the
Taylor matrixes and present the expected measurement accuracies.
The beam profile can be reconstructed by measurement of the aimuthal hit
distribution of scattered $e^+e^-$ pairs by Pair Monitor and the energy
deposit in BeamCal.

\section{Simulation}
\label{setup}
The performance of Pair Monitor and BeamCal
was studied, using the geometry of
the GLD \cite{GLD} scaled to match with 3.5 Tesla
and a simulation package
Jupiter \cite{Jupiter}. The pair background was
generated by CAIN \cite{CAIN}.
BeamCal was located at 430 cm from IP, and Pair Monitor was located
in front of BeamCal.
Solenoid field (3.5T) with the anti-DID (reversed Detector Integrated Dipole)
 was used for the magnetic field.
BeamCal is a cylinder of 18 cm radius,
which consists of 33 layers of 3.5 mm thick tungsten for
the absorber and 0.3 mm thick silicon for the sensor. 
Pair Monitor is a silicone disk of 10 cm
radius and 200 $\mathrm{\mu m}$ thickness.
There are two holes on Pair Monitor and BeamCal
whose radius are 1.0 cm and 1.8 cm for
the incoming and outgoing beams, respectively.

In this study, the nominal beam sizes are 639 nm and 5.7 nm for the
horizontal and vertical beam sizes, respectively,
as the nominal beam size.
 The center of mass energy is set to 500 GeV with the beam crossing
 angle of 14 mrad.

\section{The measurement variables}
\label{measurementVariables}
\subsection{Measurement variables of Pair Monitor}
\begin{wrapfigure}{r}{0.45\columnwidth}
\centerline{\includegraphics[width=0.4\columnwidth]{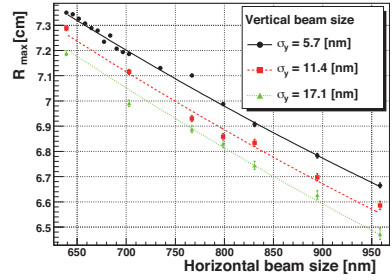}}
 \caption{$R_{shl}$ vs. $\sigma_x$ fitted with second-order polynomials.
 $R_{shl}$ decreases for larger $\sigma_x$.}
 \label{fig:rmaxVsSx}
\end{wrapfigure}
The maximum radius of hit reflects the maximum 
transverse momentum of the pairs, which in
turn is given by the electromagnetic fields of the oncoming
beam. Since the vertical beam size is much smaller than the horizontal
beam size, the maximum electromagnetic field is 
inversely proportional to the horizontal beam size.
We defined the shoulder of the radial distribution ($R_{shl}$) as the radius to contain
97.5\% of all the hits.
Figure \ref{fig:rmaxVsSx} shows $R_{shl}$
as a function of the horizontal beam size for several vertical beam
sizes. As expected,  
$R_{shl}$ decreases for larger horizontal beam size.

The azimuthal scattering angle of the pairs at the bunch crossing
would depend on the horizontal to vertical aspect ratio of the bunch,
which would then affect the 
azimuthal distribution of the hit density on Pair Monitor.
We thus studied
the distribution of the hit density as a function of the radius from the
center of the extraction beam pipe ($R$) and
the angle around it ($\phi$).
We define $N_0$ as the number of hits in $-\pi<\phi<-2.0$ radian and
$2.8<\phi<\pi$ radian for $0.6\cdot R_{shl}<R<0.8\cdot R_{shl}$. 
In order to derive the beam information from the azimuthal distribution,
we compared $N_0$ to the total number of hits ($N_{all}$).
$N_0/N_{all}$ decreases for larger horizontal and vertical beam sizes.

\begin{wrapfigure}{r}{0.45\columnwidth}
\centerline{\includegraphics[width=0.4\columnwidth]{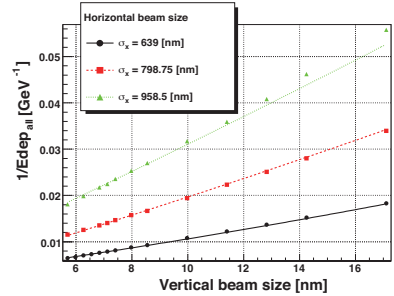}}
 \caption{$1/E_{dep_{all}}$ vs. $\sigma_y$ fitted with second-order polynomials.
 $1/E_{dep_{all}}$ increases for larger $\sigma_x$ and $\sigma_y$.}
 \label{fig:EdepVsSy}
\end{wrapfigure}
The total number of the hits on Pair Monitor, $N_{all}$, reflects the
luminosity which is inversely proportional to the vertical and horizontal beam
size.
Since the total number of the pair backgrounds are nearly
proportional to luminosity, the number of all the hits on Pair Monitor
is inversely proportional to the vertical and horizontal
beam sizes.

\subsection{Measurement variables of BeamCal}
We also investigated the measurement variables related to BeamCal.
We defined the average radius ($R_{ave}$) as follows:
\begin{eqnarray}
 R_{ave} = \frac{\sum_{i}R_i\times E_{dep_i}}{\sum_{i} E_{dep_i}},
\end{eqnarray}
where $E_{dep_i}$ means the energy deposit and $R_i$
is the radius from the center of the extraction beam pipe of the $i$-th cell.
$R_{ave}$ decreases for larger horizontal beam size similar to
$R_{shl}$ of Pair Monitor.

The total energy deposit on BeamCal, $E_{dep_{all}}$, reflects the
luminosity similar to $N_{all}$ in the case of Pair Monitor.
Figure \ref{fig:EdepVsSy} shows $1/E_{dep_{all}}$
as functions of the vertical beam size for several horizontal beam
sizes. $1/E_{dep_{all}}$ is proportional to the beam size.

\section{Reconstruction of beam sizes}
\label{reconstruction}
We reconstructed the beam sizes from the hit distribution of the pair
backgrounds on Pair Monitor and the energy deposit in BeamCal.
Five measurement variables ($R_{shl}$, $N_0/N_{all}$, $1/N_{all}$,
$R_{ave}$ and $1/E_{dep_{all}}$) are used for the
reconstruction.
These measurement variables ($m_i$, i=1, 2, $\cdots$, 5) should depend
on the beam sizes ($\sigma_x, \sigma_y$),
they can be expanded around the nominal beam sizes 
($\sigma_x^0$, $\sigma_y^0$)
 by the Taylor series.
Since these measurement variables are fitted with second-order
polynomials in this study, and $m_i$  are expanded
 up to the second-order as follows:
\begin{eqnarray}
 \label{Delta_Mi}
\Delta \vec{m} &=& 
  \vec{m}(\sigma_x, \sigma_y)
  - \vec{m}(\sigma_x^0, \sigma_y^0)
  \nonumber\\
 &=&\sum_{\alpha=x,y}
  \frac{\partial \vec{m}}{\partial \sigma_\alpha}\Delta \sigma_\alpha
 +
 \sum_{\alpha=x,y}
 \sum_{\beta=x,y}
 \frac{1}{2}\Delta \sigma_\beta\frac{\partial^2\vec{m}}
 {\partial \sigma_\alpha\partial \sigma_\beta}\Delta \sigma_\alpha
 \nonumber\\
&=&\left[A_1+\Delta\vec{\sigma}^T\cdot
	       A_2\right]\cdot\Delta\vec{\sigma},
\end{eqnarray}
where $\vec{m}=(m_1, \cdots, m_5)$,
$\Delta\vec{\sigma} =(\Delta\sigma_x, \Delta\sigma_y)$,
$\Delta \sigma_\alpha =\sigma_\alpha-\sigma_\alpha^0$,
and $A_1$ is a 5$\times$3 matrix of the first order coefficients of the Taylor expansion and $A_2$
is a tensor of the second derivative coefficients. The beam size is reconstructed by
multiplying the inverted matrix of a coefficient of $\Delta\vec{\sigma}$ in
Equation (\ref{Delta_Mi}) as follows:
\begin{eqnarray}
 \label{inverseMatrix}
 \Delta\vec{\sigma}=\left[A_1+\Delta\vec{\sigma}^T\cdot
		     A_2\right]^+\cdot\Delta\vec{m},
\end{eqnarray}
where the superscript ``+'' indicates the Moore Penrose inversion which
gives the inverse matrix of a non-square matrix $A$ as $A^+ =
(A^TA)^{-1}A^T$ \cite{fastBeamDiagnostics}.

We obtain the numerical values of the matrix ($A_1$) and
the tensor ($A_2$) of Equation (\ref{inverseMatrix}) by fitting the data by 
second-order polynomials. 
The beam sizes at IP is then reconstructed by the inverse
matrix method as shown Equation (\ref{inverseMatrix}),
where it is solved iteratively as follows \cite{fastBeamDiagnostics}:
\begin{eqnarray}
 \Delta\vec{\sigma}_n = \left[A_1 +
			 \Delta\vec{\sigma}_{n-1}^TA_2\right]^+
 \cdot \Delta\vec{m}
\end{eqnarray}
The iteration was repeated until consecutive iterations satisfied
$\left(\Delta\vec{\sigma}_n-\Delta\vec{\sigma}_{n-1}\right)
/\Delta\vec{\sigma}_n<1\%.$

\begin{wrapfigure}{r}{0.5\columnwidth}
\centerline{\includegraphics[width=0.45\columnwidth]{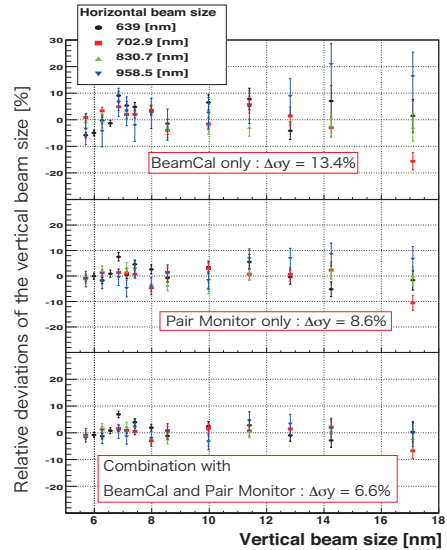}}
\caption{Relative deviations of the vertical beam size by, from the
 top, the BeamCal, Pair Monitor, and combination with the BeamCal and
 Pair Monitor.}
 \label{fig:reso}
\end{wrapfigure}

Figure \ref{fig:reso}
shows the relative deviations of the reconstructed vertical beam size
from the true beam size for 50 bunch crossings
which are measured by BeamCal, Pair Monitor, and
combination with them.
The 95\% C.L. intervals for these distributions were 13.4\%, 8.6\%, and
6.6\%, respectively. On the other hand, we obtained the 95\%
C.L. intervals of 4.4\%, 3.3\%, and 2.7\% for the horizontal beam size.
From these results, we conclude that the measurement accuracies for
the vertical and horizontal beam size are
6.6\% and 2.7\% with Pair Monitor and BeamCal.

\section{Conclusions}
\label{conclusions}
We studied a technique of beam size measurement with the Pair Monitor
and BeamCal.
The method utilizes the second-order inversion of the Taylor expansion.
Five measurement variables were used to
reconstruct the beam sizes($\sigma_x, \sigma_y$),
and the matrix elements of the expansion
were obtained
by fitting with second-order polynomials of the beam sizes.
The combined measurement with the BeamCal and Pair Monitor produces
better accuracies than that with only one of them.
The measurement accuracies of the horizontal and vertical beam sizes
for 50 bunch crossings
were found to be 2.7\% (17.3 nm) and 6.6\% (0.38 nm),
respectively.

\section*{Acknowledgments}
The authors would like to thank K.Fujii and other members of the
JLC-Software group for useful discussions and helps,
and all the member of the FCAL collaboration\cite{FCAL} for all them help.
This work is supported in part by the Creative Scientific Research Grant
(No. 18GS0202) of the Japan Society for Promotion of Science and the
JSPS Core University Program.


\begin{footnotesize}




\begin{thebibliography}{99}
\bibitem{url} Presentation: \\ 
\verb$http://ilcagenda.linearcollider.org/$\\
	\verb$contributionDisplay.py?contribId=311&sessionId=13&confId=2628$
 \bibitem{kazuto}
	 K. Ito et al.,
	 Study of Beam Profile Measurement at
	 Interaction Point in International Linear Collider,
	 arXiv:0901.4151
 \bibitem{TauchiAndYokoya}
	 T. Tauchi and K. Yokoya, Nanometer Beam-Size Measurement during
	 Collisions at Linear Colliders, KEK preprint 94-122.
 \bibitem{GLD}
	 GLD Concept Study Group: K. Abe et al. GLD Detector Outline
	 Document. physics/0607154, 2006.
 \bibitem{Jupiter}
	 Jupiter web-page:\\
	 \verb+http://acfahep.kek.jp/subg/sim/simtools/+
 \bibitem{CAIN}
	 CAIN web-page:\\
	 \verb+http://lcdev.kek.jp/~yokoya/CAIN/cain235/+
 \bibitem{fastBeamDiagnostics}
	 M. Ternick, Fast Beam Diagnostics through Beamstrahlung at
	 TESLA.
 \bibitem{FCAL}
	 FCAL collaboration web-page:
	 \verb+http://www-zeuthen.desy.de/ILC/fcal/+
\end{thebibliography}
%

\end{footnotesize}


\end{document}